# Results of a Measurement of Solar Neutrons Emitted on March 5, 2012 using a Fiber-type neutron monitor onboard the SEDA-AP attached to the ISS

**K. Koga, H. Matsumoto, O. Okudaira, T. Goka**
*Space Environment Research group, Tsukuba Space Centre, JAXA, Tsukuba, 305-8505, Japan*
E-mail: `koga.kiyokazu@jaxa.jp`

**T. Obara**
*Space Plasma Research center, Tohoku University, Sendai 980-8578, Japan*

**S. Masuda, Y. Muraki**
*Solar-Terrestrial Environment Laboratory, Nagoya University, 464-8601, Japan*

**S. Shibata**
*Engineering Science Laboratory, Chubu University, Kasugai 487-0027, Japan*

**T. Yamamoto**[*)]
*Department of Physics, Konan University, Kobe 658-8501, Japan*

Abstract

The solar neutron detector SEDA-FIB onboard the International Space Station (ISS) has detected several events from the solar direction associated with three large solar flares observed on March 5th (X1.1), 7th (X5.4), and 9th (M6.3) of 2012. In this study, we present the time profiles of those neutrons and discuss the physics that may be related to a possible acceleration scenario for ions over the solar surface. We compare our data with the dynamical pictures of the flares obtained by the ultra-violet telescope of the space-based Solar Dynamics Observatory.

1.      Introduction

When a solar flare occurs, energy stored in the sun's magnetic field is released as electromagnetic waves, including X-rays, gamma rays, radio waves, and visible light, as well as high-energy particles such as electrons and ions, which are accelerated in the process. The mechanism by which electrons are accelerated has been researched through measurement of the associated X-rays and gamma rays, and is now understood quite well. Conversely, the acceleration mechanism for ions has not been sufficiently studied because of a lack of measurement data.

To clarify the ion acceleration process, line gamma-rays have historically been studied, which are produced by the interaction between ions and the solar atmosphere (e.g., with use of the 1.02, 2.22, 4.44MeV lines). However, gamma rays are also produced by electrons; therefore, this method alone is insufficient. Therefore, we focus on the measurement of solar neutrons, which are indications of ion acceleration processes.

Because the neutrons released by nuclear reactions between the solar atmosphere and accelerated ions during solar flares are unaffected by the magnetic field both in the sun and in interplanetary space, they carry information about the energy spectrum of ions accelerated by solar flares. Neutron measurement is also important for efforts to predict solar flares in time for astronauts to be warned as a part of space weather research. Moreover, as neutrons themselves are very harmful to humankind, it is important to perceive the neutron environment around the ISS.

Many efforts toward solar neutron measurement have been accomplished since 1952. It was first pointed out by Biermann [1] that neutrons were released from the sun; the first neutron measurement in space was done by the Solar Maximum Mission (SMM) satellite on July 21, 1980 using NaI and CsI detectors [2]. On the ground, an NM64-type neutron monitor was installed on a high mountain on July 3, 1982 [3]. The measured energy of neutrons was found to be about 100 MeV, so the accelerated proton energy in the solar flare is considered to be around a few GeV.

The Space Environment Data Acquisition Equipment–Attached Payload (SEDA-AP) is the measurement unit used to observe the space environment around the International Space Station (ISS). The SEDA-AP was affixed to the Exposed Facility (EF) of "Kibo," the Japanese Experimental Module in August 2009 by the space shuttle *Endeavor*. A neutron monitor (NEM) is the one of seven instruments included in the SEDA-AP, which measures the radiation, plasma and neutral particle environments.

Since launching, we have been measuring the space environment with the use of these detectors. In this paper, we report the details of the SEDA-AP instrument, as well as the result of neutron measurements, focusing especially on the interesting event of the solar flare on March 5, 2012.

2.      Instrumentation

The NEM consists of two instruments for neutron measurement. One is a Bonner ball-type neutron detector (BBND) with an energy range from thermal to 15 MeV; the other a fiber-type neutron monitor (FIB), which can measure energies from 35 to 120 MeV. In this paper, the results of measurement by the FIB are reported.

The FIB neutron detector consists of a fiber rod of dimensions 96 mm × 6 mm × 6 mm. FIB block is built in turn about the X and Y directions using 16 layers of 16 fiber rods. Fig.1 shows the structure of the FIB. A scintillator block is surrounded by 6 scintillators for the rejection of charged particles. Light from the scintillator is guided to a 256 channel multi-anode photo-multiplier tube (PMT; Hamamatsu H4140-20), which is located in the X and Y directions, using the light guide. The high voltage value for the 256 ch PMT is about 2,000 V and the current is about 50 mA.

FIB has three observation modes: LIST, COMP, and SPECT. When the event frequency is under 15 Hz, data is acquired by LIST mode, which measures the pulse height using 8bit ADC resolution about each of the 256 channels. The COMP mode measures the pulse height with 1bit resolution from 16 Hz to 64 Hz. The SPECT mode measures only the pulse height of the dynode sum of 256 channels over 64 Hz. The amount of energy is acquired by both the dynode sum and the range of light in the FIB block. The energy resolution using dynode sum is under 39%, and that using the range of light is from 23% to 4% according to the energy range from 35 MeV to 100 MeV. In the case of the range method, as energy increases, resolution improves when the range of light is used. Neutron energy is measured using the recoiled proton track inner FIB with the relation $E_n = E_p/\cos\theta$. $E_n$ and $E_p$ are the energies of the neutron and the recoiled proton, respectively. $\theta$ is the angle between the incoming neutron (from the direction of the sun) and the recoiled proton.

A neutron detection efficiency of about 2% is obtained using Monte Carlo simulation, independent of the neutron energy [4]. The signal and noise level of each channel is evaluated for both the ground test and in space. The trigger method has two modes, one being the dynode sum of X or Y, the other being X and Y themselves. Usually when operating in the X and Y modes, the trigger level is 35 MeV, which is the minimum energy at which neutron direction is clarified by the track of the FIB. The maximum energy of this detector is about 120 MeV, which is the maximum range length of the FIB block. Beyond this energy, the track of the recoiled proton reaches the layer of the anti-counter, and these events are excluded. So, the energy range of this detector falls in the range 35 MeV–120 MeV.

The orbital period of the ISS is about 90 minutes, and one third of this period is in eclipse, during which time solar neutrons cannot be detected. Besides solar neutrons, there are local and albedo neutrons in the ISS orbit, which are produced by interactions between the ISS structure, the atmosphere and trapped radiation, or by galactic cosmic rays (GCRs), respectively. These neutrons become the background for solar neutron measurement. Typical trigger rates of the background neutrons are shown in Fig.2. The trigger rate is about a few counts per minute above the equator region and 20 counts per minute above the high latitude region; however, above the South Atlantic Anomaly (SAA) region, counts exceed about one thousand per minute, and detection of solar neutrons is not applicable.

By analyzing direction, we exclude background events from solar neutron events. Inside the solid angle within 30 degrees from the solar direction, we categorize the events as solar neutrons. Using this method, background events decreased to ~1/36.

3. Observation Results

When the SEDA was installed on the ISS, the solar activity was low and strong flares were not observed frequently. The first signal of solar neutrons was observed during the flare of March 7$^{th}$, 2011 with a GOES intensity of M3.7. With this flare, the FERMI-LAT detector also observed high-energy gamma rays being emitted continuously for several hours. The results of observations performed between 2009 and July. 2012 have been already published in our early studies [5, 6]. We will publish new results of observations for the flares with intensities of greater than M3, which were observed during 2012–2014, elsewhere. In this study, we introduce interesting flares observed from March 5$^{th}$ to 13$^{th}$ of 2012.

We analyzed the series of strong solar flares observed from March 2, 2012 to March 13, 2012. The NOAA active region AR1429 repeated strong flares. This active region moved to N17E52, and an X1.1-class flare occurred at 04:09 UT on March 5. On March 7, 00:24 UT, another X5.4-class flare was observed at N17E27, and then M6.3, M8.4 and M7.9-class flares continuously occurred on March 9, March 10, and March 13, respectively. Table I presents the result of solar neutrons in these events, whether or not the candidate solar neutrons are involved in the data. This table also contains the result of the gamma ray measurement reported by the FERMI-LAT group. The time profile of the soft X-ray intensity measured by the GOES satellite is shown in Fig. 3 and the hard X-rays measured by FERMI-GBM is shown in Fig. 4.

**Table I**

| Date | X-ray class | Location | SEDA (*neutrons*) | Fermi-LAT (*HE-γ*) | Hard X-ray peak time (*X-rays*) | Soft X-ray peak time (GOES) |
|---|---|---|---|---|---|---|
| 2012.3.2 | M3.3 | N16E83 | Weak | - | no data | 17:45 |
| 2012.3.5 | X1.1 | N17E52 | O | O | 03:42 | 04:31 |
| 2012.3.7 | X5.4 | N17E27 | O | O | 00:30 | 01:15 |
| 2012.3.9 | M6.3 | - | O | O | 03:50 | 04:01 |
| 2012.3.10 | M8.4 | - | X(eclipse) | ∼2σ | 17:55 | 17:55 |
| 2012.3.13 | M7.9 | N18W62 | O | ∼2σ | 17:23 | 17:23 |

Signals of solar neutrons were detected in the flare of March5 (X1.1), March 7 (X5.4), March 9 (M6.3), and March 13 (M7.9). The FERMI-LAT satellite also measured high-energy gamma rays accompanying these flares.

In particular in the case of March 5, the soft X-ray profile of the GOES satellite indicated one peak on the intensity curve, while the hard X-ray profile of the FERMI-GBM showed two peaks (Fig.4). However FERMI-GBM was in eclipse when the hard X-ray should have had its first peak. According to the data from the NOBEYAMA radio-heliograph, which are the ground observations with radio waves at 17 GHz and 34 GHz, the start time of the first peak is assigned at 03:32UT, and the peak time is around 03:50UT. Through comparison with these data, we analyze the data observed by SEDA-AP.

4.      Data Analysis of the Solar Flare on March 5th, 2012

The SEDA-AP was eclipsed from 03:47UT to 04:17UT, and passed through the sunshine region from 04:17UT to 05:17UT. According to the soft X-ray data of the GOES satellite, the start time of the flare was at 02:30UT, peak time was at 04:09UT, and the half-level of the declining phase was observed at 04:43UT. Concerning the hard X-ray data from FERMI-GBM, the first peak was missing and occurred when the satellite was on the sunny side of the orbit. Fig.4 suggests that the two emissions of solar neutrons are expected.

Figs. 5 and 6 show the results of analysis of the SEDA-AP data. These figures also suggest two emissions of solar neutrons. The black dot shows the observation time of the solar neutrons and the red dot shows the start time of solar neutrons according to the calculated flight time calculated on the basis of observed energy. Fig. 5 shows the results of calculation from the dynode sum value of the PMT output. The production time profile at the sun is calculated from the range of neutrons in the FIB block and the result is shown in Fig. 6. The data plotted in Fig. 6 is more accurate than that from the dynode method.

The red points form an almost-horizontal line, and this implies that the start time is the same; therefore, the two lines indicate two emissions from the times corresponding to the two peaks of hard X-ray emission. The right side of Fig. 5 shows the emission time distribution of solar neutrons calculated from the dynode energy, while the right side of Fig. 6 shows the results obtained from the range method. The horizontal axis presents the time from 04:00UT, and the vertical axis shows the counting rate per minute. These figures also suggest that neutrons were produced twice in this flare. Taking account of the error margin of the energy measurement, it is appropriate to consider the start time of neutrons to be almost the same time in each excess. The first distribution indicates a start time of 03:52UT and the second distribution starts at 04:38UT. *Therefore, protons are accelerated into high energies instantaneously when electrons are accelerated and produce hard X-rays.*

Fig 7 represents the flight direction distribution of the measurement data. The distribution forms two clusters. The ISS orbits the earth every 90 minutes, and therefore, the flight direction of neutrons reverses over the 46 minutes from 03:52UT to 04:38UT. The red line shows the solar direction from the NEM-FIB detector; this direction moves along the neutron arrival direction with time. This result also supports the claim that the excess of neutrons came from the solar direction.

5.      Data Analysis of the Solar Dynamics Observatory

In order to understand the acceleration process of ions at the solar surface, we compare SEDA data with the data obtained by the ultraviolet imager of Solar Dynamics Observatory (SDO) satellite. Fig. 8 shows the SDO data. From this picture, the following items are clarified:

(1) The coronal mass ejection (CME) started at 03:33UT. The corona rose from the solar surface at 03:36UT in the picture of the coronagraph.

(2) Loop top begun to bright of which lies from south to north. The location of the central bright point was X = −770 arcsec, Y = 370 arcsec.

(3) The signal of the NOBEYAMA heliograph was maximal at 03:55UT. According to the photograph taken at this time (03:55UT) by the SDO telescope with a wavelength of 314 nm, the x-type crossing structure between the two loops can be seen at the position (X= −770 arcsec, Y= 370 arcsec).

(4) The loop top became more intense at 04:04 UT, and hot plasma rose up. They moved in an easterly direction.

(5) A second increase in brightness was seen at 04:38UT, but the origin of light seems to have moved toward east. The bright location was X= −780 arcsec, Y= 390 arcsec.

(6) A flare shaped like the Greek character upsilon ($\Upsilon$) was observed. Two adjacent magnetic loops seemed to collide with each other at the side of both loops.

The left-hand side of Fig.8 shows the location of the first flare (03:52UT). The wavelength used in these pictures is 171 nm.

## 6.   Summary

An X-class large solar flare was observed on March 5$^{th}$, 2012. Not only high-energy neutrons but also high-energy gamma-rays were observed in association with this flare by the FERMI-LAT detector, with a statistical significance of 3.7σ. The present authors analyzed the solar neutron data and found that major emissions of neutrons occurred twice during the same flare (defined by the GOES observation) at ~03:50UT and ~04:30UT. The feature of solar neutron emission was consistent with the time profile of hard X-rays.

With the solar neutron production time having been obtained from our observations, we carefully investigated the photos taken by the ultraviolet telescope onboard the SDO satellite. Then, at the corresponding times around 03:52 UT and 04:38 UT, we found typical features of the magnetic loops: collision by the rising loop and formation of an *x-type crossing structure* and another collision between two standing magnetic loops, forming an *Upsilon-type ($\Upsilon$) collision* between the magnetic loops(right side of Fig. 8).

It has been theorized that collisions between magnetic loops cause the formation of a strong electric field, thereby accelerating ions [7], or that ions in the hot plasma are accelerated by a shock mechanism by collisions with the plasma jet [8]. The observation of SEDA-FIB seems to have found the evidence predicted by Junichi Sakai in early time with his colleague C. de Jager [9].

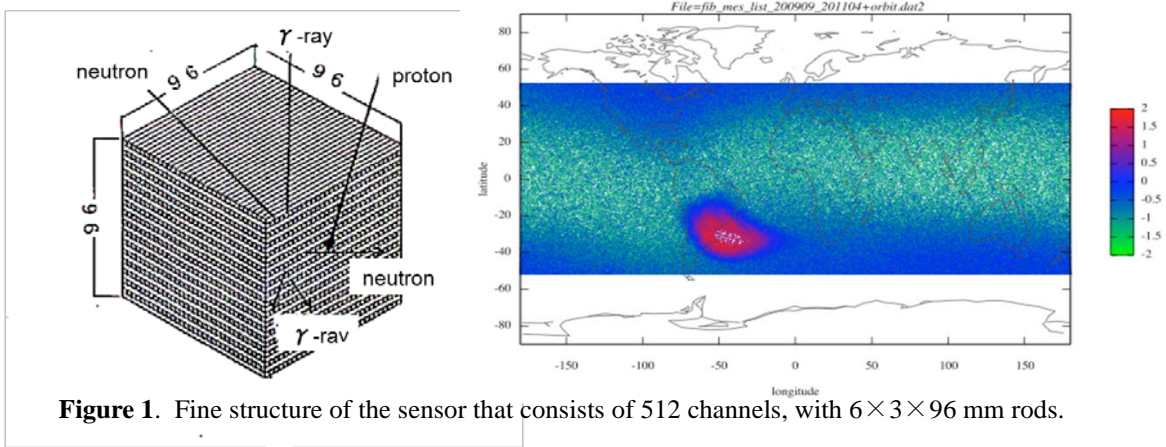

**Figure 1**. Fine structure of the sensor that consists of 512 channels, with 6×3×96 mm rods.

**Figure 2**. The map of the counting rate of SEDA-FIB. The red region corresponds to SAA.

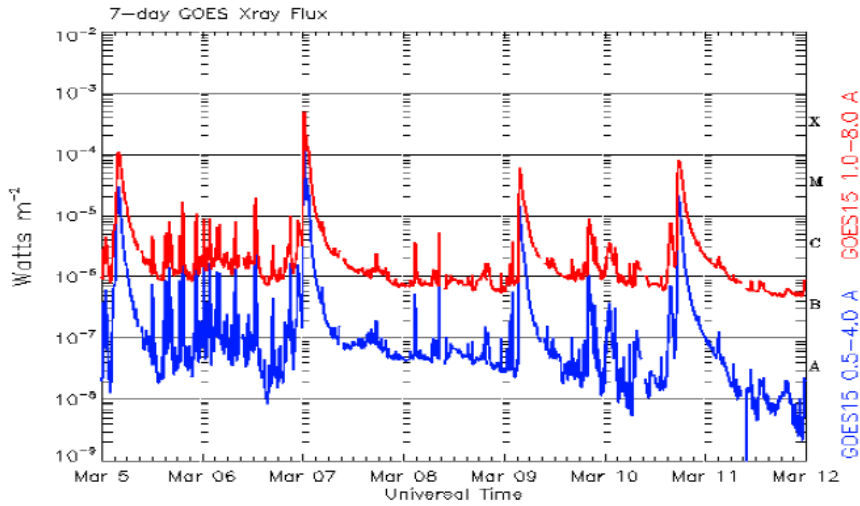

**Figure 3**. The GOES soft X-ray time profile during March 5th and March 12th of 2012.

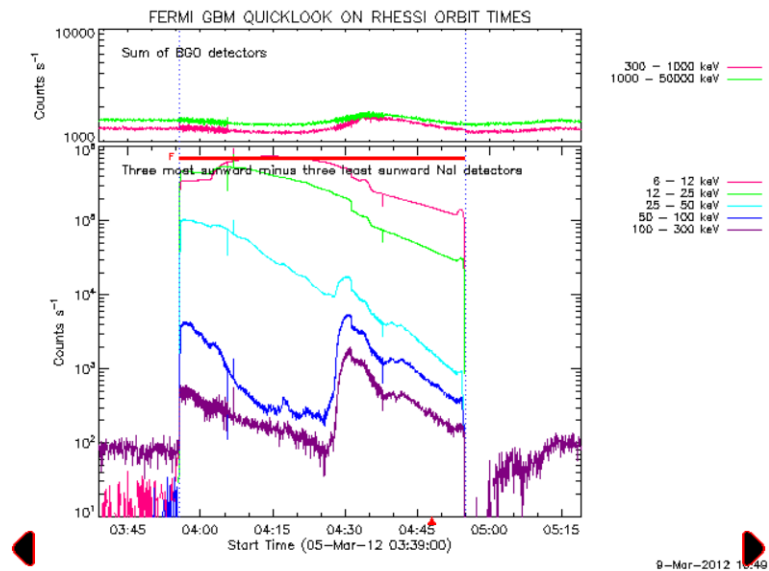

**Figure 4**. Hard X-ray intensity time profile of FERMI-GBM on March 5th, 2012 at around 04:30UT.

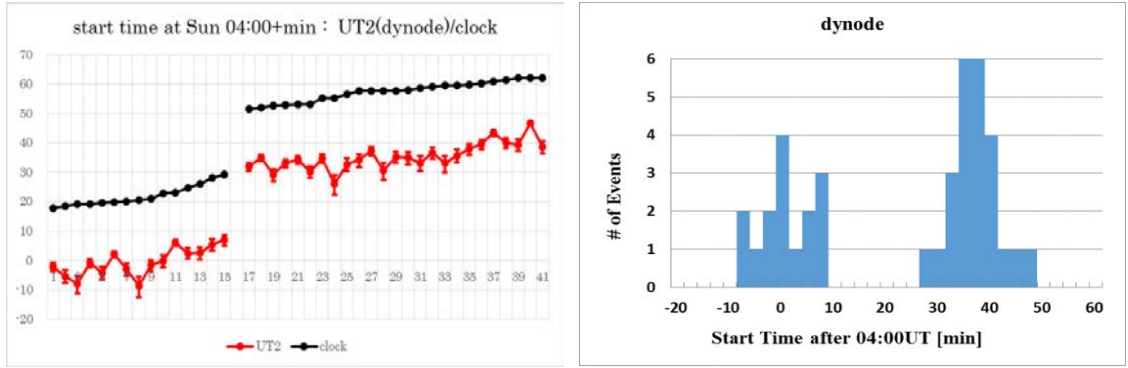

**Figure 5**. The production time distribution (red) calculated by the dynode data. The black dots correspond to the actual observation time (in minutes) from 04:00 UT.

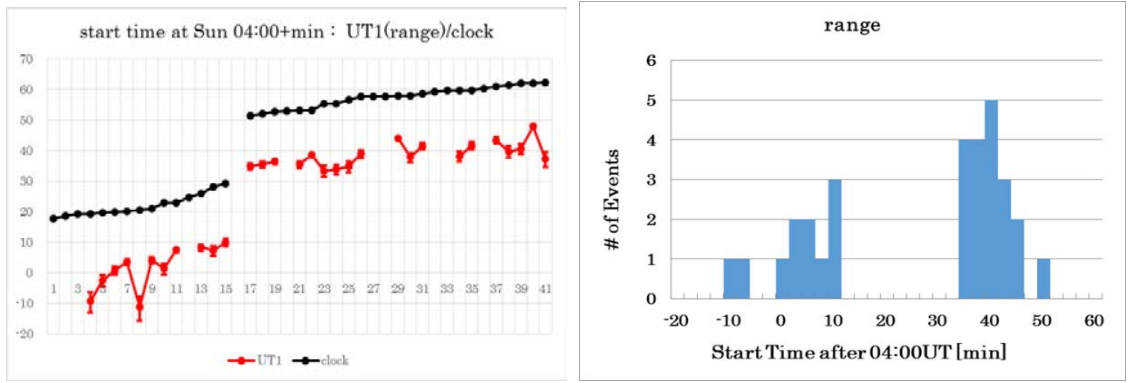

**Figure 6**. The same plot as Figure 5, but calculated by the range-counting method (better accuracy).

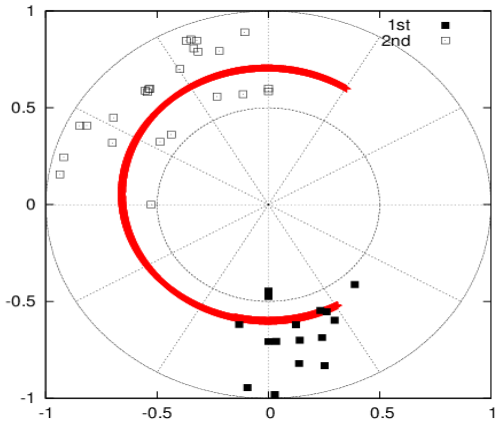

**Figure 7.** The arrival direction of solar neutrons. The red circle represents the trajectory of the Sun from 04:17UT to 05:17UT (from right side down to left top). The marks of (■) present for the first peak and the marks of (□) for the second peak of neutrons.

**Figure 8**. The SDO picture taken at 03:52 UT (left) and 04:38UT (right). The left-hand white arrow indicates the position of the x-type crossing point of the two loops. The right-hand side photo shows the ϒ type collisions. The photo was taken by using the 171nm UV telescope.

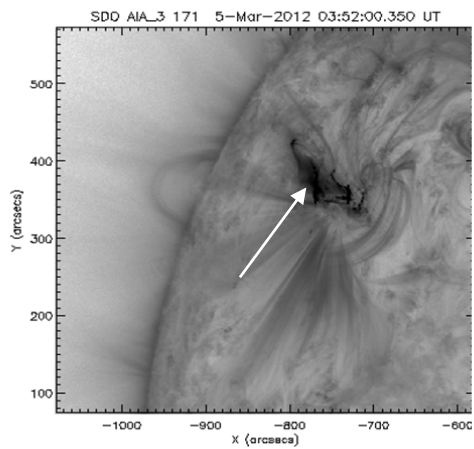
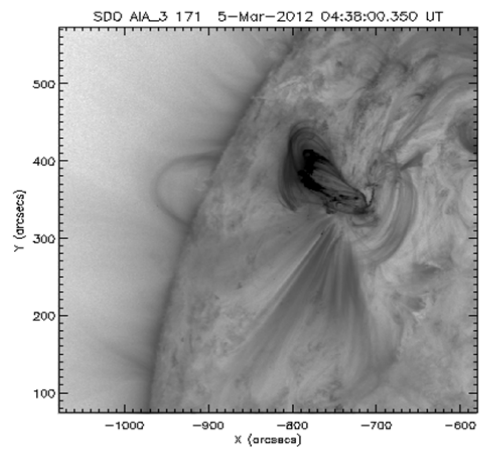